# Control of plasmonic nanoantennas by reversible metal-insulator transition

Yohannes Abate[1,2] *, Robert E. Marvel[3], Jed I. Ziegler[3], Sampath Gamage[2], Mohammad H. Javani[1,2], Mark I. Stockman[1,2], Richard F. Haglund[3,4]

1. Center for Nano-Optics (CeNO), Georgia State University, Atlanta, Georgia 30303, USA
2. Department of Physics and Astronomy, Georgia State University, Atlanta, Georgia 30303, USA
3. Interdisciplinary Materials Science Program, Vanderbilt University, Nashville, TN 37235-1406
4. Department of Physics and Astronomy, Vanderbilt University, Nashville, TN 37235-1807

* Correspondence and requests for materials should be addressed to Y.A. (email: yabate@gsu.edu)

**Nanophotonic (nanoplasmonic) structures confine, guide, and concentrate light on the nanoscale[1-3]. Advancement of nanophotonics critically depends on active nanoscale control of these phenomena[4,5]. Localized control of the insulator and metallic phases of vanadium dioxide ($VO_2$) would open up a universe of applications in nanophotonics via modulation of the local dielectric environment of nanophotonic structures allowing them to function as active devices[6-9]. Here we show dynamic reversible control of $VO_2$ insulator-to-metal transition (IMT) locally on the scale of 15 nm or less and control of nanoantennas, observed in the near-field for the first time. Using polarization-selective near-field imaging techniques, we monitor simultaneously the IMT in $VO_2$ and the change of plasmons on gold infrared nanoantennas. Structured nanodomains of the metallic $VO_2$ locally and reversibly transform infrared plasmonic dipolar antennas to monopole antennas. Fundamentally, the IMT in $VO_2$ can be triggered on femtosecond timescale to allow ultrafast nanoscale control of optical phenomena. These unique capabilities open up exciting novel applications in active nanophotonics.**

Realizing the potential of nanophotonics for signal and information processing requires control and manipulation of light at subwavelength scales. Optical energy concentration on the nanoscale is achieved on metal nanostructures due to polar electronic modes called surface plasmons (SPs)[1-3]. Photonic crystals are used for complete reflection, guiding, and confinement of light,[10] while metamaterials are used to transform light in unconventional ways, making possible such novel devices as perfect absorbers[11], circular polarizers[12], and selectively reflecting surfaces.[13] Numerous novel devices have been investigated that control the propagation of light on the nanoscale[1,14,15]. However, active nanoscale control of these phenomena in nanostructures is still a major challenge in and a bottleneck for related technologies.

In this Letter, we demonstrate unprecedented active nanoscale control of concentration of light by single plasmonic infrared antennas in the near-field. The active control of the dielectric environment by insulator-to-metal transition (IMT) in vanadium oxide ($VO_2$), dynamically transforms nanoantennas from dipole to monopole and back. We utilize the local, reversible change of refractive index of $VO_2$ that undergoes a first-order phase transition from an insulating monoclinic phase to a metallic rutile phase at approximately 69°C in bulk single crystals; the transition can also be caused by strain[16] and ultrafast light pulses[9,17]. In the IMT in polycrystalline $VO_2$ thin films, conductive nanodomains begin to nucleate and subsequently evolve to interconnect in a percolating fashion throughout the film with increasing temperature.[6] At intermediate stages of the IMT, insulating and metallic phases coexist, forming a network of high- and low-conductivity nanodomains throughout the film. Since the metallic and insulating nanodomains have substantially different refractive indices, $VO_2$ films provide for direct local control of the dielectric environment at nanometer spatial dimensions, which, in turn, can directly modulate optical responses of nanophotonic structures.

So far, the effects of the $VO_2$ IMT on plasmonic nanostructures have been studied only in the far-field[18], so that understanding and control of the near-field interaction by the $VO_2$ domains has been elusive. Here, we present an experimental study of nanoscale interactions of plasmonic structures with $VO_2$ undergoing the IMT using scattering-scanning near-field optical microscopy (s-SNOM)[19], which images local vector near-fields with minimal perturbation, indispensable for the study of nanoplasmonic phenomena[20-23]. Image formation in s-SNOM relies on the effective polarizability of tip-sample complex, allowing image contrast that is based on local dielectric environment, which is ideal for nanoscale imaging of IMT.

Our experiments were performed on an array of identical infrared plasmonic nanoantennas: gold nanorods fabricated by e-beam lithography on a 100 nm $VO_2$ film grown on a $[100]_R$ Si substrate. The dimensions of the rods (~2510 nm × 232 nm × 30 nm) were selected to be near-field resonant at mid-infrared frequencies (10.7 μm vacuum wavelength). Near-field optical images were acquired using a commercial s-SNOM system (neaspec.com). A linearly polarized $CO_2$ laser is focused on the tip–sample interface at an angle of $45^0$ to the sample surface (Figure 1b). The scattered field is acquired using phase-modulation (pseudo-heterodyne) interferometry, which yields topography, amplitude, and phase images.[21-24]

First, we investigate temperature-dependent emergence of metallic nanodomains during the IMT. The temperature was controlled by a heater, and a p-polarized (in the *yz* plane)) excitation laser was tuned to 10.7 μm, with the p-polarized detection (Methods), see Figure 1. It shows the third-harmonic (of the tip oscillation frequency) amplitude, $A_3$, images of $VO_2$ and the array of Au rods as a function of temperature. These images reveal emergence and growth of bright contrasting domains forming a local anisotropic network with increasing temperature, as the IMT progresses. While the onset of these local domains appears random, the stripes evolve by connecting to the already formed domains until the quasi-uniform metallic phase emerges at high temperatures. In the s-SNOM, higher local polarizability of the sample results in stronger near-field optical contrast, and thus these near-field images do faithfully represent the temperature-induced metallic domain formation in the $VO_2$ film.[6] Similar stripe patterns have been observed in thin epitaxial $VO_2$ films on $TiO_2$ substrates, showing preferential alignment of metallic-phase stripes due to uniaxial strain induced by the substrate.[25] Here, however, the $VO_2$ film is not locked epitaxially to the Si substrate; instead, the stripe formation appears to correlate with the direction of the nanorods. The formation and evolution of these ordered metallic stripes allows manipulation of plasmonic hot spots by nanoscale modulation of the dielectric environment.

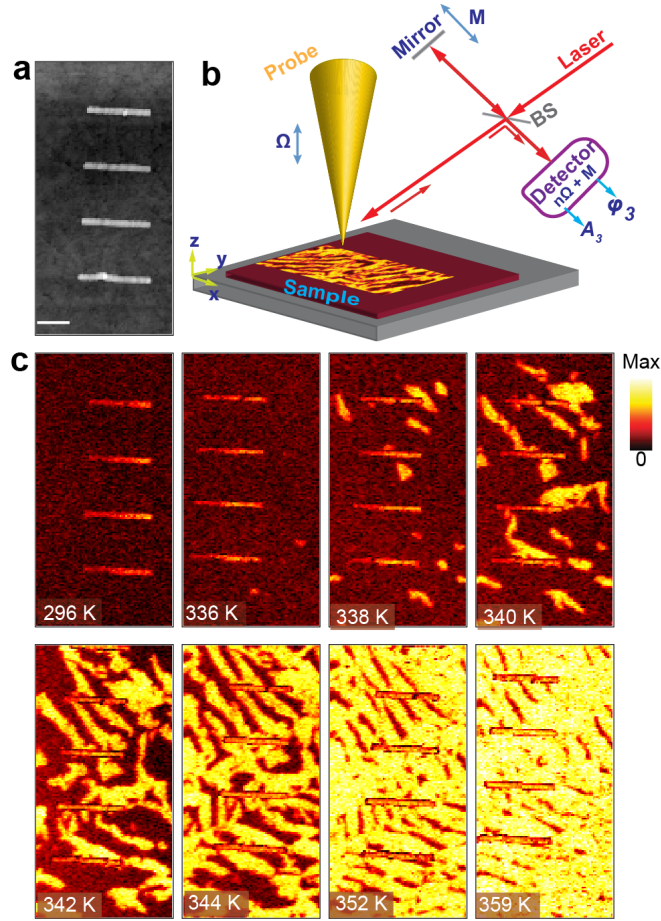

**Figure 1. Temperature dependent nanoscale near-field amplitude images of IMT emergence and progress around the infrared antennas.** a) Topography of the four antennas on $VO_2$ film. The scale bar indicates 1 μm. b) Schematic of the s-SNOM experimental setup, which allows polarization-controlled simultaneous imaging of IR plasmonic antenna modes together with the phase spatial evolution of $VO_2$ phase transition in amplitude and phase. The coordinate system is positioned so that the *y* axis is directed along nanorods and the *z*-axis is normal to the plane of the nanostructure. c) Temperature-dependent near-field amplitude images reveal IMT via forming initial metallic nanodomains, which grow and connect to stripes and quasi-uniform metallic phase.

The IMT metallic stripes appear with localized near-uniform spacing in the near-field amplitude image at an intermediate phase coexistence temperature. These are accompanied by correlating topographic modulation as clearly shown in Figure 2. Topography (Figure 2a) and near-field amplitude (Figure 2b) images are taken at excitation laser wavelength, λ=10.7 μm and temperature, T=344 K, along with the line profile sketches (Figure 2c). The topographic variation in our case is smaller (0-3 nm) and the correlation weaker compared to what was observed on a TiO$_2$ substrate (0-5 nm).[25] The correlation of the topographic line profile with the periodic stripes is due to a structural change in VO$_2$ during MIT, which results from the expansion of the rutile and the monoclinic axes.

The formation of stripe phases was noted soon after the discovery of VO$_2$. Subsequently, a link was inferred between the stripe phase and substrate strain in high-quality single crystals and epitaxial films.[25] The metallic stripe phase has been observed in thick (250 nm) VO$_2$ films grown by ion-beam assisted sputtering on TiO$_2$ [100]$_R$ surfaces.[25,26] Interestingly, the formation of the stripes is also influenced by the laser power of a near-IR pump laser (1.56 µm), close to the surface-plasmon resonance of the VO$_2$.

Here, the VO$_2$ films are thinner (100 nm), polycrystalline, and have no epitaxial relationship to the Si substrate. Nevertheless, the stripe phase appears during the transition from monoclinic to rutile, suggesting that it may occur simply because of localized, in-plane (*xy*) strains that develop at the film surface as individual grains of VO$_2$ begin to change phase, without any reference to the substrate. This possibility is further supported by the very small height of the stripes seen in the present experiment.

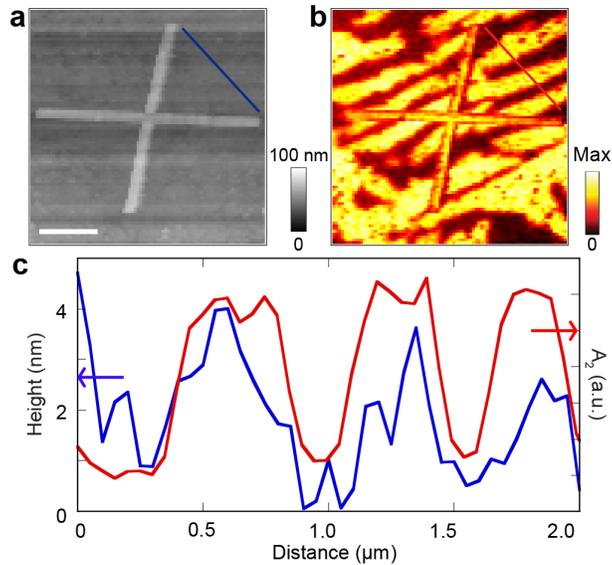

**Figure 2. Topography correlation with near-field signal.** a) Topography and b) Third harmonic near-field amplitude image showing four Au infrared antennas on VO$_2$ film. c) Topography line profile superimposed on amplitude line profile at the marked positions shown by the lines shown in (a) and (b).

To directly visualize plasmonic modes of the antennas and their interaction with the IMT of the VO$_2$ film, we implement in-plane polarization-selective excitation (s-excitation, i.e. polarized along the *y* direction) and in-plane detection (s-detection), which is referred to s/s imaging. Figure 3 shows topography, third-harmonic s/s near-field amplitudes, and phase images of the antennas on the VO$_2$ film for different temperatures. The four IR antennas are nominally identical -- see Figure 3a, which allows to compare the effects of VO$_2$ IMT on them. The amplitude images (Figure 3b-d) show bright and dark optical contrast due to the coexisting insulating (dark) and metallic (bright) phases affecting the nanoscale dielectric environment of the antennas. The metallic phase begins to form randomly with increasing temperature. As a result, portions of the antennas are located partly on the metallic and partly on the insulating phases of $VO_2$ as observed in the amplitude images (Figure 3b-d). The amplitude images allow one to see the change of nanoscale field magnitudes and the metallic phase formation. At the same time, the near-field *phase* images Figure 3e-h) are less sensitive to material contrast but allow one to follow the dipolar mode modification on each of the four antennas due to IMT. They display strong phase contrast at the rod ends. The $VO_2$ regions exhibit very weak phase contrast as shown in Figure 3e-h, which is independent of excitation or detection polarizations.

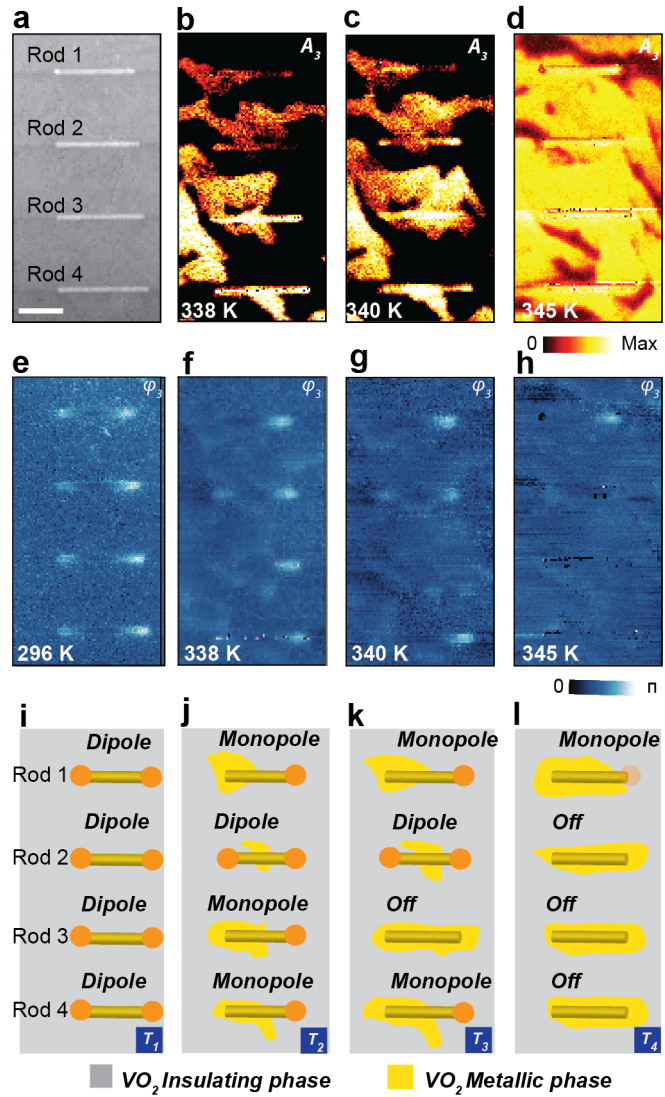

Figure 3. **Temperature-controlled IMT and antenna near-field images.** Near-field $3^{rd}$ harmonic amplitude (b-d) and phase (e-h) images. Schematics (i-l) describing experimental results of IR plasmonic antenna modes simultaneously with $VO_2$ thin film IMT domain formation and propagation.

At room temperature, all antennas display identically the expected pronounced dipolar phase contrast at their ends, as shown in Figure 3e. At higher temperatures, all antennas whose one end is situated on the metallic phase turn from dipole to monopole as evident for Rod 1 (Figure 3f-h), Rod 3 (Figure 3f), and Rod 4 (Figure 3f-g). At even higher temperature, when the amplitude image shows that most of the film is in metallic phase (Figure 3d), both dipole and monopole antenna modes of Rods 2, 3 and 4 turn off (Figure 3h) completely. An interesting case is Rod 2: despite the middle part of the rod sitting on the metallic phase (Figure 3c), it still retains its dipole (Figure 3g) since both ends are on the insulating grains. It only turns off at higher temperature when the entire antenna is situated on metal (Figure 3h). These results are interpreted in schematics shown in Figure 3i-l. This interpretation is supported by numerical calculations performed using the finite difference time-domain (FDTD) simulations (Lumerical Inc.,) shown in Figure 4, which are in excellent qualitative agreement with experiment.

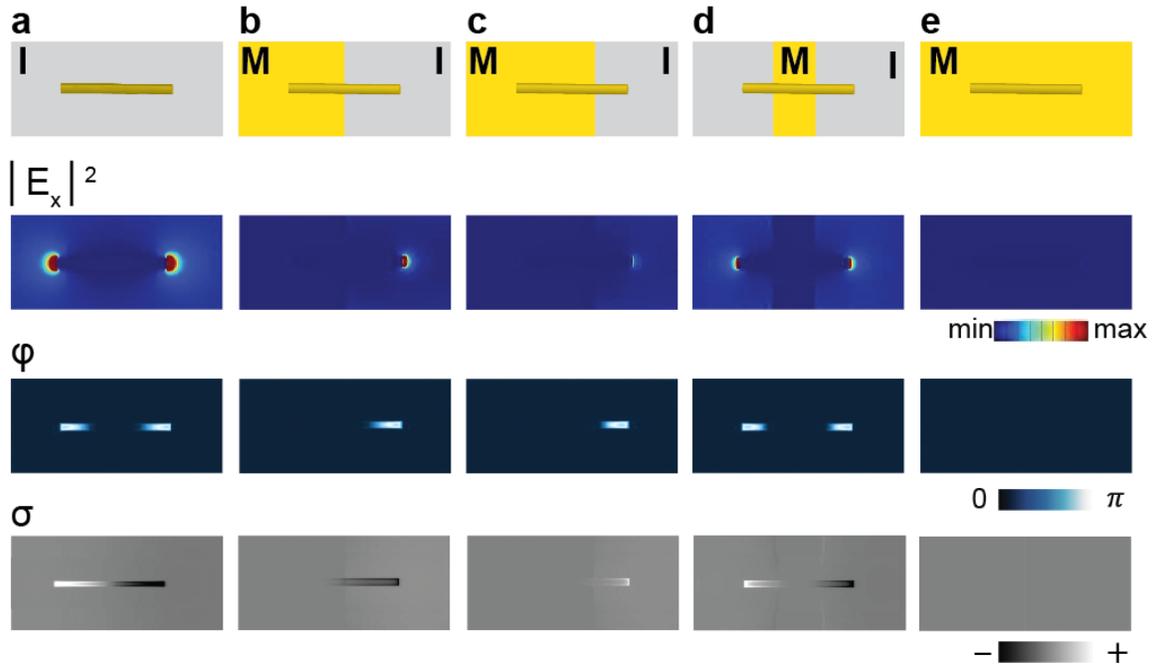

**Figure 4. Finite difference time-domain simulations.** Single antenna field intensity, phase and surface charge images of FDTD simulations. On schematic of the upper panel, I and M denote insulator and metal phases, respectively.

Further tracking of active dipole-to-monopole transformation of plasmonic antennas can be performed using s-excitation and p-detection (s/p) cross polarization selective imaging see Figure 5. Panel a displays topography, third harmonic optical near-field amplitude and phase images of an Au antenna on the $VO_2$ substrate at room temperature. The amplitude image displays a stronger optical contrast at rod ends and the phase image shows π phase difference between the rod ends. The amplitude and phase images are both signature of a dipolar mode of a plasmonic rod expected from S/P cross-polarized excitation/detection experimental method.

Figure 5b shows amplitude contrast of the Au dipolar mode simultaneously with the metallic domain at the onset of phase transition at T=341 K. Here, the amplitude optical contrast at rod ends is masked by the bright metallic domain contrast of the $VO_2$ film, and is not clearly distinguishable in the amplitude image. In contrast, the phase image (Figure 5c) distinctly discriminates the Au plasmonic rod from the metallic background of $VO_2$ film. As temperature increases (T=344 K), the metallic phase grows and a portion of one side of the rod sits on the metal and the other side sits on the insulator. The phase image between the rod ends indicates dipole (at T=296 K and 341 K) to monopole (at T=344 K) to off (at T=348 K) transformation of the nanoantenna.

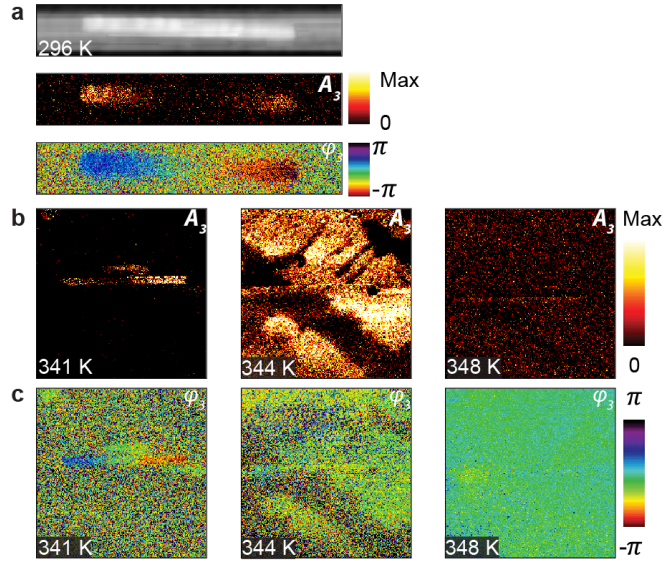

**Figure 5. Cross-polarized s/p excitation-detection imaging of plasmons and IMT of $VO_2$.** a) Topography, near-field amplitude and phase images of IR antenna at room temperature. b) near-field amplitude and c) near-field phase images of antenna on $VO_2$ film at three different temperatures.

In summary, we have shown the first experimental evidence that *near-field* local optical processes in plasmonic nanostructures can be directly and actively controlled by nanodomains in $VO_2$ film as it undergoes the IMT. Depending on the precise location of the nanoantennas with respect to metallic and insulating domains in the $VO_2$ film on the scale of 15 nm or less, the IMT reversibly transforms infrared plasmonic dipole antennas to monopole antennas or switches them off. We envision that such dynamic active control of the nanoscale interaction of light with nanostructured materials, which can potentially be ultrafast, will open up diverse applications in nanooptics and the related technologies.

## Methods

**Near-field microscopy**

The microscope is a commercial s-SNOM system (neaspec.com), which has been described in more details in detail elsewhere. A probing s or p linearly polarized $CO_2$ laser is focused on the tip–sample interface at an angle of $45^0$ from the sample surface. The scattered field is acquired using a phase modulation, or pseudoheterodyne interferometry. Suppression of the background signal is achieved by vertical tip oscillation at the mechanical resonance frequency of the cantilever ($f_0$~ 285 kHz) and demodulation of the detector signal at higher harmonics $nf_0$ (commonly $n$=2,3) of the tip resonance frequency. The combination of the scattered field from the tip and the reference beam passes through a linear polarizer which further selects the p/s polarization of the measured signal for analysis..

**Sample fabrication**

Amorphous vanadium dioxide films nominally 100 nm thick were deposited on a silicon (100) substrates by electron beam evaporation of a $VO_2$ powder precursor[27]. Annealing at 450 C in 250 mTorr oxygen crystallized the amorphous films into switching $VO_2$ with 67°C phase transition. A poly(methyl methacrylate) resist (PMMA 495 A4 from Microchem) was spun onto the $VO_2$ films before fabricating the antenna arrays by electron beam lithography with a Raith eLINE system. After development of the resist, 50 nm of gold was deposited by thermal evaporation; the remaining resist was removed by lift-off in warm acetone.

**Numerical calculations**

Experimental results are theoretically interpreted with the aid of finite difference time-domain (FDTD) simulations (Lumerical Inc., lumerical.com). For all simulations each Au rod has dimensions l=2512 nm, w= 232 nm and h= 30 nm. These dimensions were averaged from topography scan measurements. Each particle is simulated atop a $VO_2$ substrate. The optical excitation source is a mid infrared (9–11 mm) plane wave. The optical excitation source used for the simulation is a mid infrared (10.7 μm) plane wave. The simulation is performed by assuming a uniform $VO_2$ film with complex dielectric constant ε=4.9 for the monoclinic insulating phase and ε=−35+119i for the rutile metallic phase at λ=10.7 m.[28]. The simulation is performed using both s/s and s/p polarization selective excitation-detection methods.

**Acknowledgments**
YA acknowledges the major support for this work from DOD (U.S. Army Research Office) Grant No. W911NF-12-1-0076. Work of MIS was supported by Grant No. DE-SC0007043 from the Materials Sciences and Engineering Division of the Office of the Basic Energy Sciences, Office of Science, U.S. Department of Energy. REM and RFH were supported by the National Science Foundation (DMR-1207507). The samples were fabricated and characterized in facilities funded by the National Science Foundation under the American Recovery and Reinvestment Act (NSF ARI-R2 DMR-0963361).